\documentclass[conference]{IEEEtran}
\IEEEoverridecommandlockouts
\usepackage{cite}
\usepackage{amsmath,amssymb,amsfonts}
\usepackage{algorithm}
\usepackage{algpseudocode}
\usepackage{graphicx}
\usepackage{textcomp}
\usepackage{xcolor}
\usepackage{multirow}
\usepackage{makecell}
\usepackage{enumitem}
\usepackage{subcaption}
\usepackage{cleveref}
\usepackage{array}
\usepackage{adjustbox}
\usepackage{authblk}
\usepackage[left=0.764in, right=0.68in, top=0.7in, bottom=0.95in]{geometry}
\usepackage{bm}
\newcommand{\cmt}[1]{ }

\def\BibTeX{{\rm B\kern-.05em{\sc i\kern-.025em b}\kern-.08em
    T\kern-.1667em\lower.7ex\hbox{E}\kern-.125emX}}
\begin{document}

\title{A Fog-based Smart Agriculture System to Detect Animal Intrusion \vspace{-0.2in}}
% {\footnotesize \textsuperscript{*}Note: Sub-titles are not captured in Xplore and
% should not be used}
% \thanks{Identify applicable funding agency here. If none, delete this.}

\author{Jinpeng Miao$^{1}$, Dasari Rajasekhar$^{2}$, Shivakant Mishra$^{1}$, Sanjeet Kumar Nayak$^{2}$, Ramanarayan Yadav$^{3}$}
\affil{$^{1}$University of Colorado Boulder, USA, $^{2}$IIITDM Kancheepuram, India, $^{3}$IITRAM, Ahmedabad, India}

% \author{\IEEEauthorblockN{1\textsuperscript{st} Given Name Surname}
% \IEEEauthorblockA{\textit{dept. name of organization (of Aff.)} \\
% \textit{name of organization (of Aff.)}\\
% City, Country \\
% email address or ORCID}
% \and
% \IEEEauthorblockN{2\textsuperscript{nd} Given Name Surname}
% \IEEEauthorblockA{\textit{dept. name of organization (of Aff.)} \\
% \textit{name of organization (of Aff.)}\\
% City, Country \\
% email address or ORCID}
% \and
% \IEEEauthorblockN{3\textsuperscript{rd} Given Name Surname}
% \IEEEauthorblockA{\textit{dept. name of organization (of Aff.)} \\
% \textit{name of organization (of Aff.)}\\
% City, Country \\
% email address or ORCID}
% \and
% \IEEEauthorblockN{4\textsuperscript{th} Given Name Surname}
% \IEEEauthorblockA{\textit{dept. name of organization (of Aff.)} \\
% \textit{name of organization (of Aff.)}\\
% City, Country \\
% email address or ORCID}
% \and
% \IEEEauthorblockN{5\textsuperscript{th} Given Name Surname}
% \IEEEauthorblockA{\textit{dept. name of organization (of Aff.)} \\
% \textit{name of organization (of Aff.)}\\
% City, Country \\
% email address or ORCID}
% \and
% \IEEEauthorblockN{6\textsuperscript{th} Given Name Surname}
% \IEEEauthorblockA{\textit{dept. name of organization (of Aff.)} \\
% \textit{name of organization (of Aff.)}\\
% City, Country \\
% email address or ORCID}
% }

\maketitle

\begin{abstract}
Smart agriculture is one of the most promising areas where IoT-enabled technologies have the potential to substantially improve the quality and quantity of the crops and reduce the associated operational cost. However, building a smart agriculture system presents several challenges, including high latency and bandwidth consumption associated with cloud computing, frequent Internet disconnections in rural areas, and the need to keep costs low for farmers. This paper presents an end-to-end, fog-based smart agriculture infrastructure that incorporates edge computing and LoRa-based communication to address these challenges. Our system is deployed to transform traditional agriculture land of rural areas into smart agriculture. We address the top concern of farmers - animals intruding - by proposing a solution that detects animal intrusion using low-cost PIR sensors, cameras, and computer vision. In particular, we propose three different sensor layouts and a novel algorithm for predicting animals' future locations. Our system can detect animals before they intrude into the field, identify them, predict their future locations, and alert farmers in a timely manner. Our experiments show that the system can effectively and quickly detect animal intrusions while maintaining a much lower cost than current state-of-the-art systems.
\end{abstract}

\begin{IEEEkeywords}
smart agriculture, animal intrusion detection, LoRa, fog computing\footnote{We use the term `fog computing' and `edge computing' interchangeably throughout the paper, same as `fog server' and `edge server'. }
\end{IEEEkeywords}

\section{Introduction}
%Smart agriculture is the application of IoT technology to traditional agriculture and the use of sensors and IoT platforms to automatically and intelligently control agricultural production. 
Smart agriculture applies modern information technology, integrates big data, mobile Internet, cloud computing and IoT technologies relying on various sensing nodes to achieve precise tracking, monitoring, automating and analyzing operations. At present, cloud-based infrastructures are being utilized to support various smart agriculture applications and data processing. Data from smart sensors in the agricultural field is transmitted to the cloud over the Internet, and then stored and processed in the cloud for decision making. While cloud-based infrastructures certainly offer enormous processing power and storage capacity, there are two key limitations that need to be addressed when used in the context of smart agriculture~\cite{b1}: (i) Sensor data transmitted over the Internet requires continuous Internet connectivity, consumes high bandwidth and incurs delays. (ii) Since IoT devices must transmit large volumes of data to the cloud for storage and processing, the energy of battery-powered IoT devices is quickly drained. These limitations make cloud-based infrastructure ill-suited for smart agriculture. To address these limitations, we propose a LoRa-enabled, fog-based smart agriculture infrastructure that reduces the quantity of data transferred to the cloud and enable latency-sensitive services delivered just in time. 

After conducting a survey with the farmers to understand the key issues they are facing that could be addressed by smart agriculture, animal intrusion in the field becomes the most concerning one. Farms are usually located in rural areas, close to nature. This makes animal intrusion a major issue for farm owners who must deal with the mess and damage these animals can cause. Compared to some other smart services such as smart irrigation, crop quality monitoring and pest extermination, animal intrusion detection is more difficult because of its uncertainty, uncontrollability, unpredictability. Animals may eat crops and stroll around the field at any time, resulting in a significant production loss. This necessitates more time costs to recover from the damage as well as greater financial security to cover the costs associated with damages.  

In this paper, we propose an end-to-end, LoRa (\textbf{Lo}ng \textbf{Ra}nge)-enabled, fog-based infrastructure for smart agriculture along with a new strategy to detect animal intrusion. We are committed to helping farmers detect and locate animal invasions as quickly as possible. We firstly introduce how the low-power, low-bandwidth and long-range features of LoRa are utilized to transform traditional agriculture lands in rural areas into smart agriculture system. We then present the design and implementation of a microservice-based edge server that provides important, latency-sensitive services to the farmers and enables operation in a disconnected Internet environment. To enable the fastest detection of animal intrusion, we explore several sensor placement strategies and design an algorithm which is able to locate invasive animals and predict future locations. Finally, we evaluate the performance of our proposed system and compare with the existing, state-of-the-arts frameworks in terms of cost, latency and distance. 

This paper makes the following contributions:
\begin{itemize}
\item Adoption of LoRa protocol effectively addresses the limitations of intermittent Internet connectivity and high latency of cloud-based infrastructure.
\item A microservice-based architecture at the edge to enable latency-sensitive services delivered just in time.
\item Propose three sensor layouts and an algorithm that accurately predicts the future locations of animals.
\item Comprehensive analysis and comparison of the layouts through experiments. 
\item Rigorous evaluation and discussion on the accuracy of the algorithm and the practicality of the system.
\end{itemize}

\section{Background}

%In this section we present an overview of the background that led us to the proposed solution. We start by presenting an overview of communication techniques including LoRa and LoRaWAN. Then we introduce IoT devices we use in this work and some edge computing related technologies, such as containerization and microservices, which are the technologies that we leverage in our solution. 

\subsection{LoRa and LoRaWAN Protocol}
% A graph shows how LoRa and LoRaWAN works
LoRa is an ultra-long-distance wireless transmission technology based on spread spectrum technology~\cite{b24,b25}. LoRaWAN is a set of communication protocol and system architecture designed for long-distance communication network~\cite{b23,b25}. Lora has a great advantage in handling co-channel interference. It solves the problem of not being able to take into account long distance, anti-interference and low power consumption at the same time. Compared with other communication technologies, LoRa's ultra-low cost, high sensitivity, ultra-low power consumption, strong anti-interference ability, low bandwidth consumption, and long transmission distance make it ideal for this project.

\subsection{IoT Devices}

\subsubsection{Arduino}
a microcontroller-based open source hardware platform. Arduino Mega is Arduino development board based on the ATmega. It is cheap and easy-to-use features make it widely used in practical IoT projects.  

\subsubsection{Multi-channel LoRaWAN GPS concentrator}
a high-performance multi-channel transmitter/receiver designed to receive multiple LoRa packets simultaneously. It is intended to provide a robust connection between a central wireless data concentrator and a large number of wireless endpoints over a considerable range of distances.  

\subsubsection{Passive infrared (PIR) sensor}
an electronic sensor that measures infrared (IR) light radiating from objects in its field of view. 
%As shown in Figure~\ref{pir}, 
The characteristics of this sensor include the angle of detection ($\alpha$) and the maximum detectable distance ($d$) with the detection range calculated as a cone with $h$ as the diameter of the circle as the base. It is small, cheap,  power efficient, easy to use and durable. Therefore, they are typically used for security and automatic lighting related purposes.

%\begin{figure}[ht]
%\vspace{-0.1in}
%\centering
%\includegraphics[width=0.19\textwidth]{figures/pir.png}
%\caption{Field of View of PIR Sensor.}
%\label{pir}
%\vspace{-0.2in}
%\end{figure}

\subsubsection{All-day camera}
a camera that is able to detect invasive animals both day and night, which requires a built-in motorized IR-cut filter so that it can switch in and out automatically based on light condition. The filter will be turned off with the purpose that only visible light during the daylight, and IR sensitivity during the night with IR LEDs on. 

\subsection{Fog Computing}
\subsubsection{Containerization}
is a software deployment process that bundles the application's code with all the files and libraries the application needs to run on any infrastructure~\cite{b22}. By virtualizing the operating system kernel, this technology enables user-space software instances to be divided into multiple independent units that run in the kernel as opposed to a single instance. This particular software instance is referred to as a \textbf{container}, a software package that provides the complete runtime environment for an application. With containerization, people can create individual packages or containers that can run on all types of devices and operating systems. Containerization is lightweight, portable, scalable, fault-tolerant, agile, and saves hardware resources.

\subsubsection{Microservices}
is a type of software architecture that builds complicated programs using modularity and small functional units that are each focused on a particular responsibility and function. Microservices architecture makes applications easier to scale and faster to develop. Compared to monolithic architecture, microservices architecture is agile, scalable, easy to deploy, technically free, code-reusable, and resilient~\cite{b21}. 

% By using microservices and containerization technologies, we have the flexibility to add more intelligent services to our already established system in the future.
\section{Proposed System}
% A general architecture graph includes everything

\subsection{System Architecture}

\begin{figure}[ht]
\vspace{-0.1in}
\centering
\includegraphics[width=0.48\textwidth]{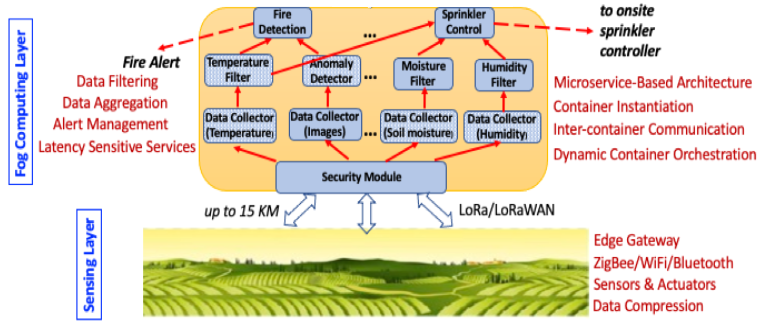}
\caption{Proposed system architecture.}
\label{general_sys}
\vspace{-0.1in}
\end{figure}

The architecture of the proposed microservice-based fog enabled infrastructure for smart agriculture is shown in Figure~\ref{general_sys}. It consists of two layers: sensing layer and fog computing layer, which are linked by cross-layer upstream and downstream communication for data and control information flows~\cite{b28}. 

The {\it sensing layer} is comprised of the sensors and actuators deployed across the agricultural field to periodically sense the physical parameters of interest such as air temperature, air humidity, soil temperature and moisture at various depths, wind speed and rainfall. %In a typical scenario, agriculture fields tend to be at remote locations with low-bandwidth and poor Internet connectivity. Furthermore, farmers need to remotely monitor large agriculture fields spanning distances of 5 km to 10 km, or even more. 
To address the challenge of poor Internet connectivity, we have adopted a LoRa and LoRaWAN enabled communication system due to their support for low power, wide area networking designed to wirelessly connect limited energy operated IoT devices to an edge server at a distance of 1-2 km.
The {\it fog layer}
is comprised of one or more servers, and provides an administrative control of the entire IoT infrastructure of the agricultural field. It addresses the limitations of intermittent Internet connectivity, high latency and high network bandwidth consumption of cloud-based infrastructure. 
The fog nodes 
%collect data from the sensors deployed in the field, perform key data cleaning, filtering, aggregation and fusion tasks, and 
execute latency sensitive services, such as animal intrusion detection. To facilitate a flexible architecture that may utilize existing container-based support for various ML/AI services, we have structured the fog layer as a microservice architecture. In this architecture, an application is composed from a collection of loosely-coupled microservices, where each microservice is fine-grained and the associated protocols are lightweight. 

\subsection{Animal Intrusion Detection}

\begin{figure}[ht]
\vspace{-0.1in}
\centering
\includegraphics[width=0.4\textwidth]{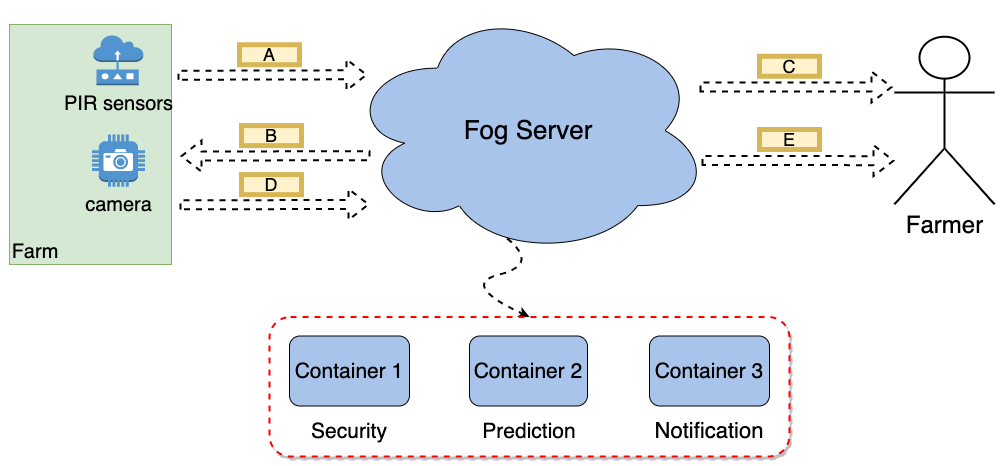}
\caption{System architecture for animal intrusion detection.}
\label{sys}
\vspace{-0.1in}
\end{figure}

In view of the serious problems caused by animal intrusion to farmers, our goal is to automatically detect animal intrusions, identify animals, and inform the farmer(s) in a timely manner about the intrusion. This work is performed using two types of sensors: a PIR sensor for detecting any motion in its field of view and an all-day camera sensor attached to the Raspberry Pi for capturing images that will be processed to identify animals. To meet the low-latency requirement, the scheduling mechanism and the prediction algorithm are implemented in the fog layer, while the object detection is done on the Raspberry Pi. This is because the low bandwidth of LoRa cannot support the transmission of large-sized images. As shown in Figure~\ref{sys}, there are three microservices: 1) The Security module passes the sensor data it receives from authenticated sensors to the appropriate Prediction container; 2) The Prediction container runs localization and prediction algorithms on this data as well as recorded data, and then sends the predicted position at a future time to the camera; 3) The Notification module notifies the farmer via messages once animals are detected. 

Corresponding to the process marked with capital letters in Figure~\ref{sys}, the steps are described as follows: 
\begin{enumerate}[label=\underline{\textbf{\Alph*}}:]
\item Animal movement is detected by PIR sensors and the data is transmitted to the server using LoRa.
\item A container on the edge server predicts the location of the animal at a future time based on input from multiple sensors and sends this location to the Raspberry Pi that operates a camera on the field.
\item The edge server sends a ``possible animal invasion" alert to the farmer.
\item The Raspberry Pi instructs the camera to rotate in the direction of the predicted position and take a picture. The Raspberry Pi then runs an animal detection algorithm on this image and sends the results to the edge server.
\item The edge server sends a reliable alert to the farmer, who can then decide an actuation.
\end{enumerate}

% \subsection{From Sensors to Fog}
% The sensing layer consists of end nodes whose task is to sense the agriculture field parameters and communicate them to the fog layer using LoRa communication protocol. The end node is responsible for sending sensor data to the fog node. 
% in the format of ``side-sensor-timestamp" 
 % or receiving the actions/commands from fog nodes

\subsection{Sensor Layouts}
\label{sec:layout}
\begin{figure}[ht]
\vspace{-0.1in}
\centering
\includegraphics[width=0.3\textwidth]{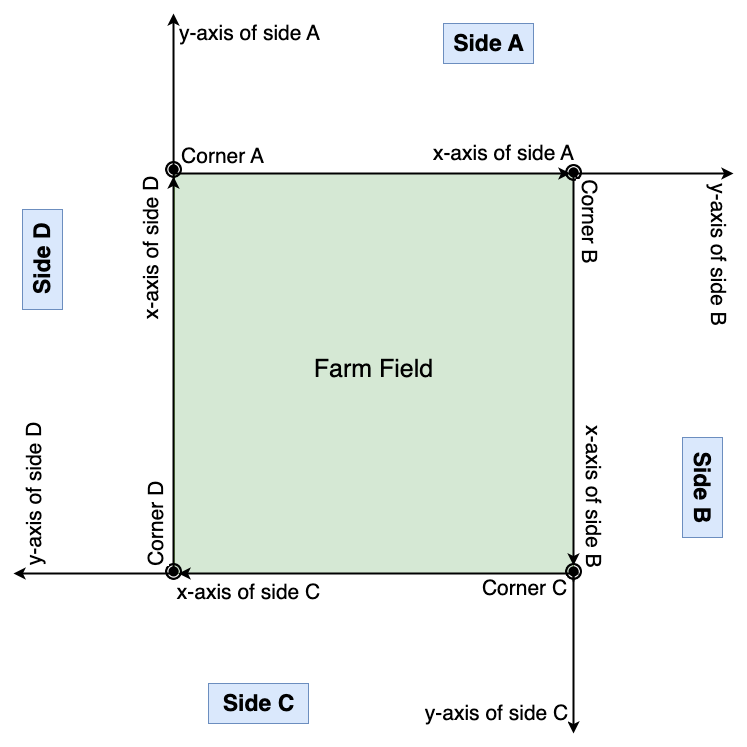}
\caption{Virtual coordinate systems built upon the farm.}
\label{coordinate}
\vspace{-0.2in}
\end{figure}

In this section we introduce three different sensor layouts which could yield different localization and prediction accuracy. There will be some independent coverage areas between the sensors as well as overlapping areas. We establish a virtual coordinate system upon the farm field (as shown in Figure~\ref{coordinate}) and use the center coordinates of a small area to represent this area. In other words, when sensors detect an animal, the animal's location is regarded as a point (marked with $R1$, $R2$, \ldots in~\Cref{layoutA,layoutB,layoutC}) rather than a range, which is convenient for us to design algorithms to predict animals' position. In order to describe the specific location of the animal to the farmer, we define four corners and four sides. Below we describe the three sensor layouts in detail.
% Through the detection results of the sensor, the position and movement trajectory of the animal can be tracked and predicted.

\subsubsection{Layout A - Vertical}
\begin{figure}[ht]
\vspace{-0.1in}
\centering
\includegraphics[width=0.4\textwidth]{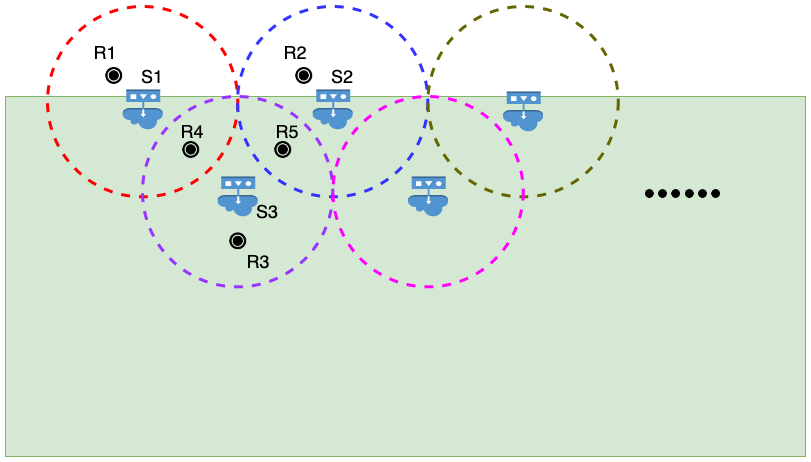}
\caption{Layout A: Vertical Placement}
\label{layoutA}
\vspace{-0.1in}
\end{figure}

We place all sensors at a height of $d$ meters above the ground and project them vertically downward, with the coverage area of each sensor being a circle of diameter $h$. This produces a coverage area consisting of many circles. As shown in Figure~\ref{layoutA}, we put two rows of interlocking and overlapping sensors at the boundary of the field. This not only increases the coverage, but the overlapping areas allow for relatively fine-grained segmentation of the area to improve the accuracy of localization and prediction. The increase in budget associated with an additional row of sensors is well worth it compared to more accurately catching animal intrusions and thus preventing damage to the farm. But the shortcoming of this layout is that the farthest detectable location is too close to the farm boundary, only $h/2$, which leads to a greater chance of animal damage to the farm.

\subsubsection{Layout B - Horizontal}
\begin{figure}[ht]
\vspace{-0.1in}
\centering
\includegraphics[width=0.4\textwidth]{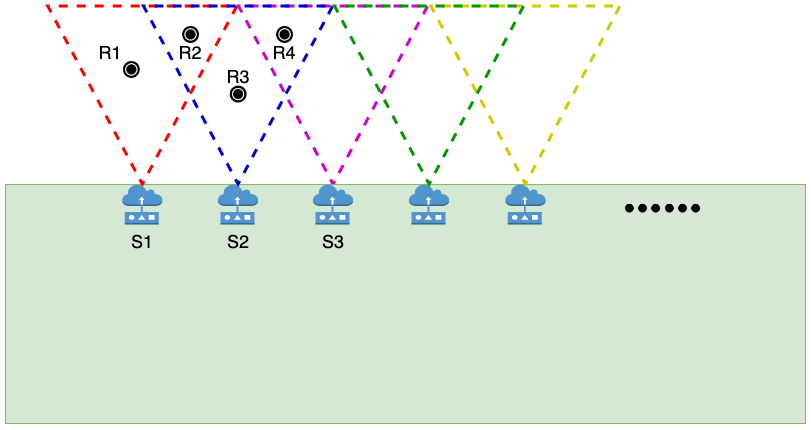}
\caption{Layout B: Horizontal Placement}
\label{layoutB}
\vspace{-0.1in}
\end{figure}

In contrast to Layout A, all sensors are placed on the ground and horizontally projecting towards the farm's exterior, with each sensor covering an isosceles triangle with $h$ as the base and $d$ as the height, resulting in a coverage area of many triangles. As shown in Figure~\ref{layoutB}, we put one row of interlocking and overlapping sensors at the boundary. This also has the same advantages as Layout A, i.e., increased coverage and fine-grained area segmentation to improve localization and prediction accuracy.
Moreover, it overcomes the limitations of Layout A by extending the farthest detectable distance, thus improving protection.

\subsubsection{Layout C - Hybrid}
\begin{figure}[ht]
\vspace{-0.1in}
\centering
\includegraphics[width=0.4\textwidth]{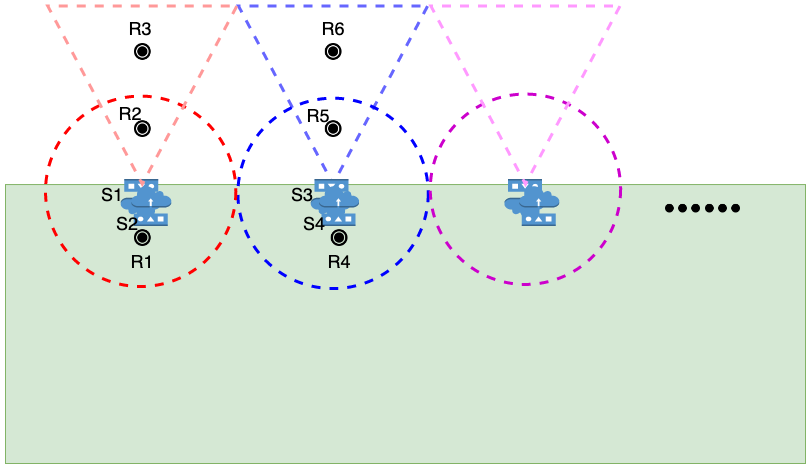}
\caption{Layout C: Hybrid Placement}
\label{layoutC}
\vspace{-0.1in}
\end{figure}

With vertical and horizontal placement, it was natural to explore a hybrid placement. We still place two rows of sensors, one vertically along the boundary and the other projected horizontally outward at the same location, as shown in Figure~\ref{layoutC}. This layout provides good coverage, fine-grained area segmentation, and a far-reaching detectable location. However, uncovered middle areas can lead to inaccurate or even outrageous predictions. Additionally, this layout has a calculated minimal coverage area.

In addition to these three layouts, we also considered other layouts that required fewer sensors. However, they were ruled out due to their limited coverage, which limits their prediction accuracy, and their short sensing distance to the boundary, which make them unsuitable for fast-moving animals. 
%Given the importance of the layout and the low cost of sensors, we ultimately choose the above three layouts.

\subsection{Proposed Algorithm}
We propose and deploy an algorithm (as shown in Algorithm~\ref{alg:predict}) on the fog server to predict the future location of intrusive animals based on the previous readings returned by the sensors. Whenever a sensor detects an animal, it sends the data back to the container running the algorithm on the fog server in the format of ``\#side-\#sensor-timestamp”. Each set of data received by the container is combined with the previous record to make a prediction. The ``set of data" here may have two scenarios: one is the data read back from a non-overlapping coverage area; the other is the animal is in the overlapping area covered by multiple sensors. In this case there will be multiple sensors with similar timestamps to send back data and the server needs to make the final prediction after receiving data from all these sensors. We use a 0.1 second ``tolerance time difference" to define this similar timestamp. In addition, we define another threshold of 120 seconds as the minimum time interval between animal intrusions, i.e., if the server does not receive new data within 120 seconds, the next data received is considered to be a new animal intrusion.

We define a mapping of sensor numbers and position coordinates in the algorithm. The server first converts the sensor number in the received data into a coordinate and combines the previous set of coordinates to calculate the distance and direction, and then to calculate the average speed of the animal's movement with the timestamp difference. With the direction and speed, the next position of the animal can be predicted under the assumption that the animal will move in the same direction with the same speed for a short period of time. This period is the sum of the time it takes for the sensor to return data, the time it takes for the algorithm to make the prediction, the time it takes for the instruction to be passed from the server to the Raspberry Pi, and the time it takes for the camera to rotate to point to the predicted position.

The direction and speed of animal movement are not stable, but the constant detection and updating of position information by the sensors, the fast transmission of LoRa, and the high speed calculation of the system can make the predicted deviation be calibrated quickly and continuously. The field of view of the camera can also provide a certain degree of tolerance. Taken together, our proposed algorithm is expected to effectively and accurately locate and predict the location of animals. Next, we evaluate and verify the adequacy of the algorithm through experiments.

\vspace{-0.05in}
% pseudo code for the algorithm
\begin{algorithm}
\caption{Algorithm to predict animal locations}
\label{alg:predict}
\begin{algorithmic}[1]
\Statex \textbf{Input:} side number $side$, sensor number $sensor$, and timestamp $t_{cur}$
\Statex \textbf{Output:} A coordinate of predicted animal position \{$x_{predict}, y_{predict}$\}

\State $x_{prev}$   \Comment{x value of previous location}
\State $y_{prev}$   \Comment{y value of previous location}
\State $t_{prev}$   \Comment{Timestamp of previous reading}
\State $time\_threshold \leftarrow 120$
\State $time\_tolerance \leftarrow 0.1$
\State $latency$    \Comment{Time required from detection to camera pointing to the predicted position}
\State $pos\_mapping \leftarrow \{sensor : \{x : y\}\}$

\Function{Predict}{$side$, $sensor$, $t_{cur}$}
    \State $x_{cur} \gets pos\_mapping[sensor][x]$
    \State $y_{cur} \gets pos\_mapping[sensor][y]$
    \If{$t_{cur} - t_{prev} > timing\_threshold$}
        \State \textit{Do nothing}
    \ElsIf{$t_{cur} - t_{prev} < time\_tolerance$}
        \State \textit{Wait until all data received}
    \Else
        \State $dist \gets \sqrt{(x_{cur} - x_{prev})^2 + (y_{cur} - y_{prev})^2}$
        \State $speed \gets dist / (t_{cur} - t_{prev})$
        \State $\theta \gets \operatorname{arctangent }(y_{cur} - y_{prev}, x_{cur} - x_{prev})$
        \State $d \gets latency * speed $
        \State $x_{predict} \gets x_{cur} + d * math.cos(\theta)$
        \State $y_{predict} \gets y_{cur} + d * math.sin(\theta)$
        \State \Return ${x_{predict}, y_{predict}}$
    \EndIf
    \State $x_{prev} \gets x_{cur}$
    \State $y_{prev} \gets y_{cur}$
    \State $t_{prev} \gets t_{cur}$
\EndFunction
\While {true}
    \State \Call{Predict}{$side$, $sensor$, $t_{cur}$}
\EndWhile
\end{algorithmic}
\end{algorithm}
\section{Evaluation}
All experiments presented in this paper using the parameters shown in Table~\ref{table:sys_config}. To evaluate our work, we constructed an end-to-end LoRa communication system, deployed the three sensor layouts proposed in Section~\ref{sec:layout}, and gathered sensing data by moving along various trajectories. We then implemented our proposed algorithm to analyze the collected data.

% Configuration of device
\begin{table}[htbp]
\caption{System Configuration}
\vspace{-0.1in}
\begin{center}
\begin{tabular}{|c|c|}
 \hline
 \textbf{Component Name} & \textbf{Specifications} \\
  \hline
 Arduino Mega &  \thead{256 KB Flash Memory, 8KB SRAM,\\ 4KB EEPROM, 16 MHz Clock Speed} \\
 \hline
 Raspberry Pi &  \thead{Quad core Cortex-A72 (ARM v8) 64-bit \\ 8GB RAM, 1.5GHz Clock Speed} \\
 \hline
 PIR Sensor  &  \thead{Detection range $d$ is 7 meters; \\ Detection distance $h$ is 5 meters} \\
  \hline
Camera & \thead{Resolution 2592×1944, Optical Size 6.35mm, \\ Focal Length 2.25mm, FOV 130°(D) 105°(H)} \\
 \hline
 Edge Server  & \thead{3.1 GHz Dual-Core Intel Core i5,\\ 8 GB RAM,  256GB Disk} \\
  \hline
\end{tabular}
\label{table:sys_config}
\end{center}
\vspace{-0.1in}
\end{table}

\subsection{Experiments and Results}
\subsubsection{Lora Transmission}
\begin{figure}[ht]
\vspace{-0.1in}
\centering
\includegraphics[width=0.37\textwidth]{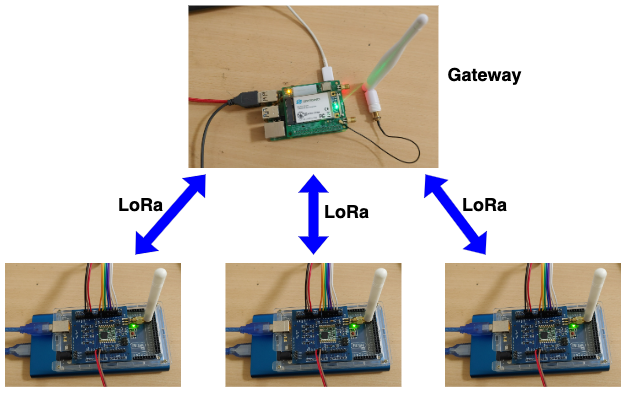}
\caption{LoRa communication using multiple end nodes.}
\vspace{-0.1in}
\label{connect}
\end{figure}

We build an end node which consists of PIR sensors, one Arduino Mega microcontroller, LoRa Hat with Antenna. LoRa hat is built using LoRa SX1276 IC. To experiment the scheduling capability of fog node, we connected three end nodes with one LoRa enabled gateway (Raspberry Pi with PG1302 LoRaWAN Concentrator) as shown in Figure~\ref{connect}. The end nodes are scheduled in a round robin fashion by fog node to avoid interference of data during simultaneous communication by the three end nodes.

Table~\ref{table:lora_latency} shows the experimental results obtained while communicating data from end node to the fog gateway. We varied the heights of sender and receiver and also the distance between them and checked the delay in LoRa communication. We performed these experiments in an environment where various objects and building were present (these obstacles cause channel attenuation and thus affect the received signal strength). Data that is communicated between sender and receiver is 16 bytes. It can be observed that as we increase distance between sender and receiver, latency increases. Latency can be reduced further by placing receiver (fog node) at proper height.

\begin{table}[htbp]
\caption{LoRa experiment result}
\vspace{-0.1in}
\begin{center}
\begin{tabular}{|c|c|}
 \hline
 \textbf{Sender / Receiver Node (Height, Distance)} & \textbf{Latency(s)} \\
  \hline
 Sender: 4ft, Receiver: 4ft, Sender ~ Receiver: 500m & 4.0 \\
 \hline
 Sender: 4ft, Receiver: 35ft, Sender ~ Receiver: 350m & 0.7 \\
 \hline
 Sender: 35ft, Receiver: 200ft, Sender ~ Receiver: 500m & 1.1 \\
  \hline
Sender: 4ft, Receiver: 200ft, Sender ~ Receiver: 2000m & 0.9 \\
  \hline
\end{tabular}
\label{table:lora_latency}
\end{center}
\vspace{-0.1in}
\end{table}

\subsubsection{Sensor Placement}
\begin{figure}
  \vspace{-0.1in}
  \centering
  \begin{subfigure}[b]{0.12\textwidth}
    \includegraphics[width=0.9\textwidth]{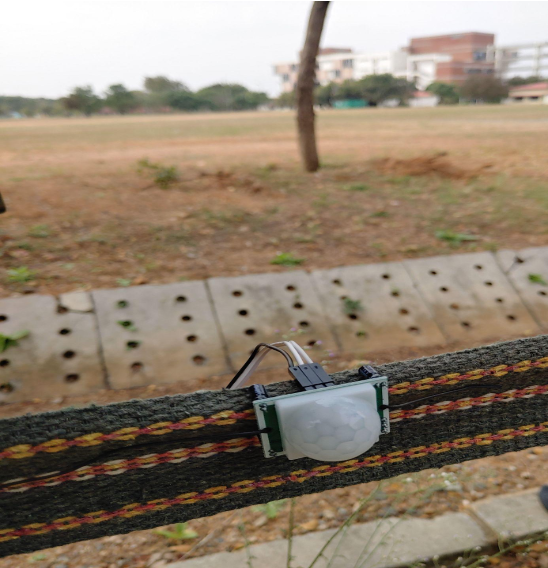}
    \caption{Placement 1}
    \label{p1}
  \end{subfigure}
  \begin{subfigure}[b]{0.12\textwidth}
    \includegraphics[width=0.9\textwidth]{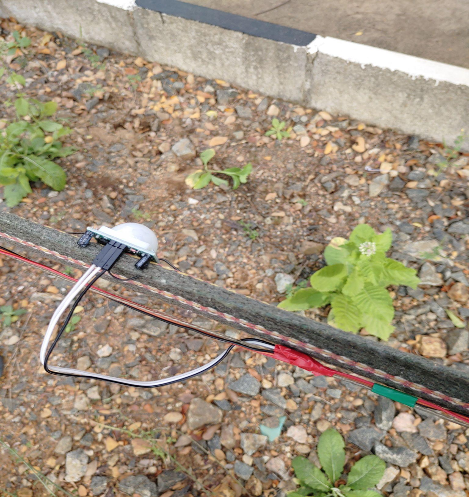}
    \caption{Placement 2}
    \label{p2}
  \end{subfigure}
  \begin{subfigure}[b]{0.12\textwidth}
    \includegraphics[width=0.9\textwidth]{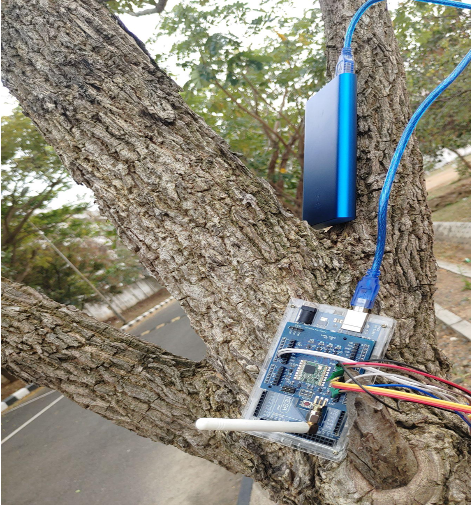}
    \caption{Placement 3}
    \label{p3}
  \end{subfigure}
    \begin{subfigure}[b]{0.12\textwidth}
    \includegraphics[width=0.9\textwidth]{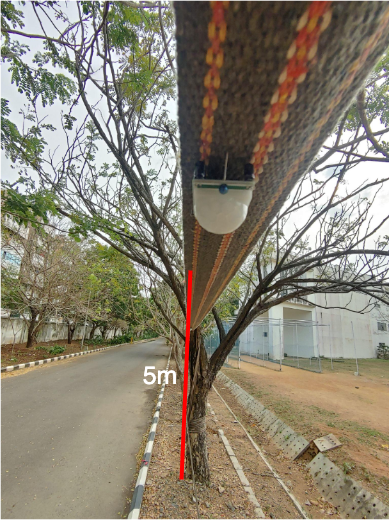}
    \caption{Layout A}
    \label{la}
  \end{subfigure}
    \begin{subfigure}[b]{0.12\textwidth}
    \includegraphics[width=0.9\textwidth]{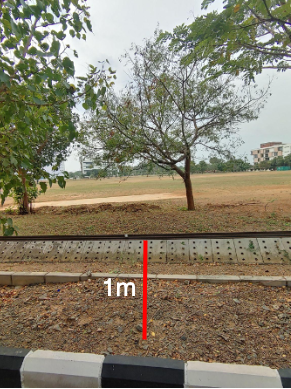}
    \caption{Layout B}
    \label{lb}
  \end{subfigure}
    \begin{subfigure}[b]{0.12\textwidth}
    \includegraphics[width=0.9\textwidth]{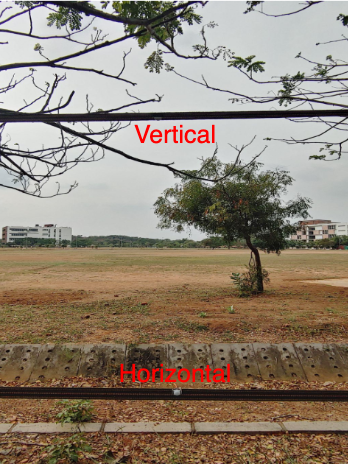}
    \caption{Layout C}
    \label{lc}
  \end{subfigure}
  \caption{Sensor Placement}
  \label{sensor_place}
  \vspace{-0.1in}
\end{figure}

To detect animal presence in a 25 by 25 meters field, we utilize PIR sensors, whose specifications are detailed in Table~\ref{table:sys_config}. Figure~\ref{sensor_place} shows the specific experimental deployment for the sensor layouts. The PIR sensors were fixed to a strip and placed around the perimeter of the field to ensure complete coverage. Vcc and GND common wires were connected to each sensor, while the output pin was connected separately with a different wire for each sensor to the Arduino board. We used an Arduino ATMega2560 along with a LoRa hat using 
 LoRa SX1276 IC powered by a lithium-ion battery to send data to the Gateway for processing and decision-making.

\begin{figure}[ht]
\vspace{-0.1in}
\centering
\includegraphics[width=0.3\textwidth]{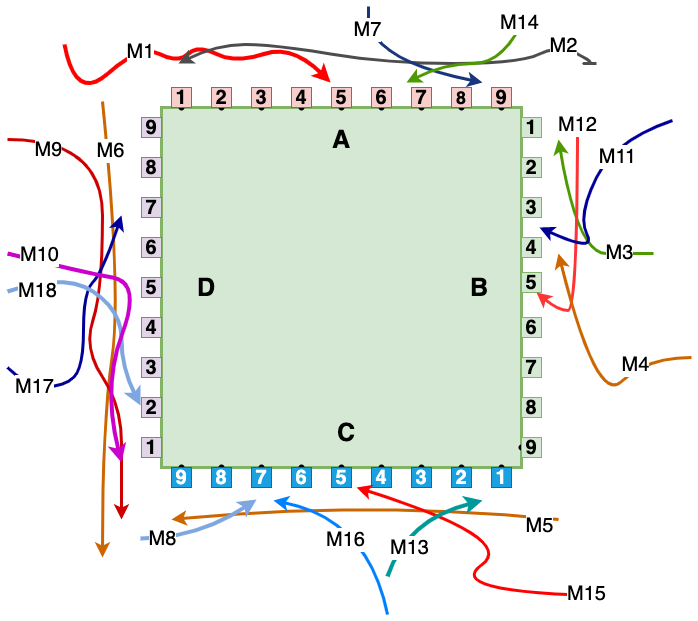}
\vspace{-0.1in}
\caption{Movement Trajectories}
\label{traj}
\vspace{-0.1in}
\end{figure}

For layout A, the distance between two sensors was set to 5 meters, and the strip was placed 5 meters above the ground, facing towards the ground to allow for detecting activity in the area immediately below the strip, within a 5 meter radius around each sensor. In layout B, PIR sensors were horizontally oriented and placed on a strip, with a uniform distance of 2.5 meters between each sensor, providing consistent coverage across the field. The strip was placed at a height of 1.5 meters above the ground and faced outward towards the field. Layout C utilized a hybrid approach, with sensors arranged in two strips: one positioned horizontally 1.5 meters above ground level and the other arranged vertically towards the ground at a height of 5 meters. The sensors were spaced evenly at 5 meter intervals on each strip. We changed different speeds, directions, and trajectories to simulate 18 different movements (M1 - M18 in Figure~\ref{traj}) of animals to evaluate the accuracy and effectiveness of our tracking algorithm. For each of these 18 movements, our system sensed and transmitted PIR sensor values to the edge server for multiple locations depending on which sensors detected movements. For example, Table~\ref{tab:movementdata} shows the the location values received for Movement M2. This location data collected from 18 different movements forms the ground truth for our evaluation.

\begin{table}[htbp]
\caption{M2 location values: side – coordinate(x, y)}
\vspace{-0.1in}
\begin{center}
\begin{tabular}{|c|c|c|c|}
 \hline
  & \textbf{Layout A}  & \textbf{Layout B} & \textbf{Layout C}  \\
 \hline
 1  & A - (20.00, 1.5) &  A - (20.00, 3.5) &  A - (20.00, 5.50) \\
 \hline
 2  &  A - (15.00, 1.5) &  A - (15.00, 3.5) &  A - (15.00, 5.50) \\
 \hline
 3  &  A - (15.00, 1.5) &  A - (15.00, 3.5) &  A - (15.00, 5.50) \\
 \hline
 4  &  A - (10.00, 1.5) &  A - (10.00, 3.5) &  A - (10.00, 5.50)  \\
 \hline
 5  &  A - (5.00, 1.5) &  A - (5.00, 3.5) &  A - (5.00, 5.50)  \\
 \hline
\end{tabular}
\label{tab:movementdata}
\end{center}
\vspace{-0.1in}
\end{table}

\subsubsection{Position Prediction}
\begin{figure*}[t]
\vspace{-0.1in}
\includegraphics[width=0.9\textwidth]{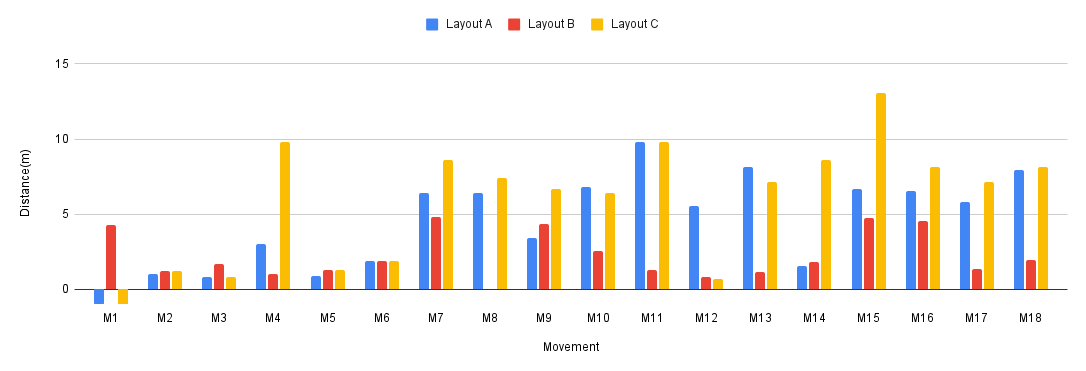}
\centering
\vspace{-0.15in}
\caption{Average distance offset between predicted and actual positions for different types of movements in the three layouts.}
\label{compare}
\vspace{-0.1in}
\end{figure*}

We use a laptop computer as fog server in our experiments. The laptop was placed high so as to increase the transmission speed with end nodes according to Table~\ref{table:lora_latency}. The prediction algorithm runs continuously waiting for data. To evaluate our algorithm, we make use of our ground truth data, wherein
the container extracts three sets of location data from a movement, uses the first two sets of data to predict the location for the time corresponding to the third set of data, and then compares this predicted location with the actual location to assess the accuracy of the prediction. One measure of accuracy we use is the {\it distance offset}, which is the distance between the predicted location and the actual location. We measured distance offsets for all movements for which we have at least three location values. For movements such as M2 (Table~\ref{tab:movementdata}) for which we have more than three location values, we measured distance offset for each triplet of location values resulting in 10 distance offsets measured. Figure~\ref{compare} shows the average distance offset of each movement. 

As we can see, the average distance offset is relatively low (less than 5 meters) for most movements and layouts. We observe that Layout B shows relatively small distance offsets in most of the movement tests, although in M1, it has a higher offset in prediction than the other two layouts. However, in M8, Layout B does not have sufficient readings for the algorithm to make predictions due to the presence of some blind triangles near the boundary where the sensor cannot detect the animal once it moves there. Layout C produces the largest distance offsets in most of the tests due to the presence of many blind areas inside the coverage area, which prevented the animal from being detected quickly and continuously, resulting in more inaccurate predictions. Layout A performs moderately and without data loss, which is due to its continuous and extensive coverage area. Based on our experiments and the analysis in Section 2, we recommend Layout B as the optimal sensor deployment method.

\subsubsection{Animal Detection}
To identify the intrusive animals, we connect a camera to a Raspberry Pi to take pictures of the animals (area where the predicted location is) and identify them using computer vision algorithms. The specifications of the Raspberry Pi are shown in Table~\ref{table:sys_config}. We experimented with several popular CNN pre-trained models to test their speed of processing images. We first train these models on top of a laptop and then imported the trained models into the Raspberry Pi. These event-driven models are continuously running on the Raspberry Pi waiting for images to be taken. 
The average detection time (based on 20 runs for each image) is summarized in Table~\ref{table:img_latency}. As we can see, these models take 2 to 5 seconds to identify an animal, with MobileNet having the best performance with an average time of 1.64 seconds.

\begin{table}[htbp]
\caption{Animal Detection Experiment Results}
\vspace{-0.1in}
\begin{center}
\begin{tabular}{|c|c|}
 \hline
 \textbf{CNN Pre-Trained Model} & \textbf{Latency(s)} \\
  \hline
 VGG16 & 2.27 \\
 \hline
 ResNet50 & 3.75 \\
 \hline
 ResNet50V2 & 3.34 \\
  \hline
 InceptionV3 & 4.75 \\
  \hline
 MobileNet & 1.64 \\
  \hline
 MobileNetV2 & 2.74 \\
  \hline
 EfficientNetB0 & 5.07 \\
  \hline
\end{tabular}
\label{table:img_latency}
\end{center}
\vspace{-0.1in}
\end{table}

%\subsection{Layout Comparison}
%From Figure~\ref{compare}, we can observe that Layout B shows relatively small distance offsets in most of the movement tests, although in M1, it has a higher offset in prediction than the other two layouts. However, in M8, Layout B does not have sufficient readings for the algorithm to make predictions due to the presence of some blind triangles near the boundary where the sensor cannot detect the animal once it moves there. Layout C produces the largest distance offsets in most of the tests due to the presence of many blind areas inside the coverage area, which prevented the animal from being detected quickly and continuously, resulting in more inaccurate predictions. Layout A performs moderately and without data loss, which is due to its continuous and extensive coverage area. Based on our experiments and the analysis in Section 2, we recommend Layout B as the optimal sensor deployment method.

\subsection{Prediction Accuracy}
\label{sec:accuracy}
\begin{figure}[ht]
\vspace{-0.1in}
\centering
\includegraphics[width=0.48\textwidth]{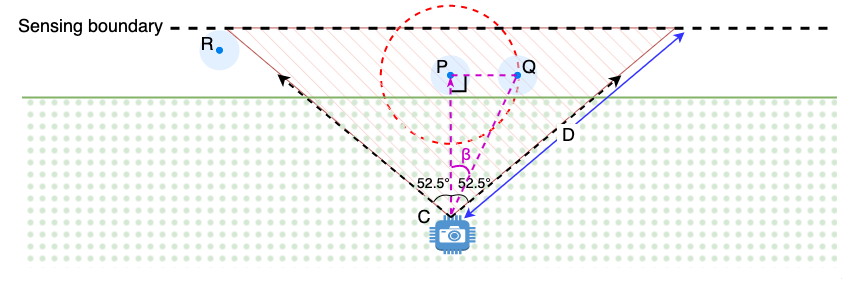}
\caption{Camera placement}
\label{camera_placement}
\vspace{-0.1in}
\end{figure}

The goal of predicting location is to be able to rotate the camera in a direction where the animal is expected to be. Using the distance offset statistics, we can determine the accuracy of the algorithm by combining the distance between the predicted animal position and the camera placement. As illustrated in Figure~\ref{camera_placement}, the predicted location is represented by point $P$, and the camera (point $C$) is positioned within the boundary to point towards the predicted position. The camera has a horizontal field of view of 105 degrees, as shown in Table~\ref{table:sys_config}, and the red shading indicates the current range that the camera can cover. If the actual animal position falls within the red shading, represented by point $Q$, the prediction is considered accurate. Conversely, if the actual animal position falls outside the red shading, represented by point $R$, the prediction is considered incorrect. Since we only have the distance between the predicted location and the actual location, without knowing their relative positions with respect to the camera, $Q$ could be any place on the red circle which is centered at point $P$. We make the assumption that the angle formed by the edge $PQ$ and the edge $CP$ at point $P$ is a right angle, so that angle $\beta$ is the maximum value. In this way, the accuracy of the prediction is the most conservative value.

Based on this validation method, we calculate the prediction accuracy of the three layouts at different distances between the predicted animal position and the camera placement, as shown in the Table~\ref{table:camera_accuracy}. The table shows again that layout B is the best layout solution. For layout B, a placement distance of 5 meters can achieve a very accurate prediction. The farther the distance, the wider the coverage, and the higher the accuracy. Nevertheless, we must also consider that increasing the distance results in lower image quality of the animal, which makes animal identification more difficult. We will discuss this further in Section~\ref{sec:discuss}.

\begin{table}[htbp]
\caption{Camera placement and prediction accuracy.}
\vspace{-0.1in}
\begin{center}
\begin{tabular}{|c|c|c|c|}
 \hline
   \multirow{2}{2.6cm}{\centering{Distance between camera and animal(m)}} & \multicolumn{3}{c|}{Accuracy(\%)} \\ \cline{2-4}
   & Layout A  & Layout B & Layout C  \\
 \hline
 5  & 66.67\% & 94.44\% & 38.89\%  \\
 \hline
 10  & 100\% & 94.44\% & 94.44\%  \\
 \hline
 15  & 100\% & 94.44\% &  100\%  \\
 \hline
\end{tabular}
\label{table:camera_accuracy}
\end{center}
\vspace{-0.1in}
\end{table}

\subsection{System Performance and Cost}
\subsubsection{System Performance}
Based on our experiments, we estimate the total time required to achieve animal intrusion detection with the current system configuration (as shown in Table~\ref{table:sys_config}), which is summarized in Table~\ref{table:latencies}. It takes about 19.11 to 28.61 seconds for the farmer to get clear information about the intrusion, including the location and type of the animal. Furthermore, the farmer will receive successive messages to calibrate the animal's location until the danger is removed.

\vspace{-0.1in}
\begin{table}[htbp]
\caption{System Latency}
\vspace{-0.1in}
\begin{center}
\begin{tabular}{|c|c|}
 \hline
 \textbf{Step} & \textbf{Latency(s)} \\
  \hline
 Transmission of 3 sets of data via Lora & $3 \sim 9$ \\
 \hline
 Latency between 3 readings & 10 \\
 \hline
 Prediction with proposed algorithm &  0.01 \\
 \hline
 Instruction sent to camera via LoRa & 1 \\
\hline
 Camera rotation, image capture and processing & $4 \sim 7$ \\
 \hline
 Results sent back to fog server via LoRa & 1 \\
 \hline
 Alert sent to farmer via LTE & $0.1 \sim 0.6$ \cite{b27} \\
 \hline
 In total & $19.11 \sim 28.61$ \\
 \hline
\end{tabular}
\label{table:latencies}
\end{center}
\vspace{-0.1in}
\end{table}

\subsubsection{System Cost} 
To illustrate the expenses incurred during our experiment in the $25 \times 25$ $m^2$ field, we have compiled a detailed cost analysis of all the devices used, which is presented in Table~\ref{table:device_list}. With a total cost of 823 US dollars, our system is a cost-effective solution for monitoring animal intrusions. As the size of the farm increases, the cost will inevitably rise, but the advantage is that additional expensive equipment, such as fog server, is not required.

% A list of devices used in the project along with their costs
\vspace{-0.1in}
\begin{table}[htbp]
\caption{System Cost}
\vspace{-0.15in}
\begin{center}
\begin{tabular}{|c|c|}
 \hline
 \textbf{Device Name} & \textbf{Cost(US Dollars)} \\
  \hline
 Arduino Shield for LoRa & 28\$/each x 2 =  56\$  \\
 \hline
 Raspberry Pi 4  & 95\$ \\
 \hline
 GPS Concentrator & 120\$ \\
 \hline
 PIR Sensor & 0.75\$/each x 36 = 27\$ \\
 \hline
 Camera & 25\$ \\
 \hline
 Laptop & 500\$ \\
 \hline
 In Total & 823\$ \\
 \hline
\end{tabular}
\label{table:device_list}
\end{center}
\vspace{-0.1in}
\end{table}

\section{Discussion}
\label{sec:discuss}
% \subsection{Camera Placement}
In Section~\ref{sec:accuracy}, we assessed the accuracy of prediction based on the presence or absence of animals in the picture taken. However, to fully evaluate the performance of the system, we must also consider the ability to identify the pictured animals. If the animal image is not clear in the picture, it will be difficult to identify. This depends on two factors: the number of pixels that the animal occupies in the image and the pixel requirements of animal recognition algorithms listed in Table~\ref{table:crop_pixels}~\cite{b26}. Assuming the animal size is approximately 2 meters long and 1.5 meters high, we calculate the number of pixels occupied by the animal at different distances between the camera and the animal based on the camera parameters (as shown in Table~\ref{table:sys_config}) and present the results in Table~\ref{table:cnn_pixels}.

By comparing these two tables, we can confirm that these widely used models listed can successfully identify animals when the distance between the animal and the camera is 40 meters, and we can still use the very effective GoogLeNet and SqueezeNet1\_1 when the distance is 80 meters. Therefore, to ensure both a large camera coverage to improve the quality of animal imaging and the tolerance of prediction errors, we need to control the camera placement and maximum rotation angle. Specifically, we must ensure that the maximum distance between the camera and the intersection of the coverage boundary and the sensing boundary (i.e., $D$ in Figure~\ref{camera_placement}) does not exceed 80 meters, with 40 meters being the optimal distance. This allows us to adopt MobileNet, which has been shown to achieve the best performance in Table~\ref{table:img_latency}.

\begin{table}
    \centering
    \caption{Pixels an animal occupies at different distances}
    \label{table:cnn_pixels}
    \resizebox{\columnwidth}{!}{
    \begin{tabular}{|c|c|c|c|c|c|c|c|c|}
        \hline
         \textbf{Distance(m)} & 10  & 20  & 30  & 40  & 50  & 60  & 70  & 80\\
        \hline
        \textbf{Pixels} & 199x151 & 99x76 & 66x50 & 50x38 & 40x30 & 33x25 & 28x22 & 25x19\\
        \hline
    \end{tabular}
    }
\end{table}

\begin{table}
    \centering
    \caption{Minimum pixel requirement for CNN models}
    \label{table:crop_pixels}
    \resizebox{\columnwidth}{!}{
    \begin{tabular}{|c|c|c|c|c|c|}
         \hline
         \textbf{Model} & GoogLeNet & SqueezeNet1\_1 & DenseNet201 & VGG16/19 & MobileNet \\
          \hline
         \textbf{Minimum Pixels} &  15x15 &  17x17 & 29x29 & 32x32 & 32x32 \\
         \hline
    \end{tabular}
    }
\end{table}
\section{Related Work}
With the proliferation of smart agriculture, many related systems have been proposed. However, most suffer from safety hazards, excessive cost and resource consumption, reliance on Internet connectivity, or poor performance. Given the number of systems in the literature, we focus on the latest intelligent agricultural systems and animal intrusion detection strategies.

Devaraj \emph{et.al.} suggest using traditional electric fence, which shock animals that cross the boundary~\cite{b2}. While effective and easy to install, it needs constant power supply and regular maintenance. It becomes ineffective during power outages, whereas our system remains operational. Notably, non-intelligent electric fences may harm animals and people. In~\cite{b6}, authors analyze why traditional methods such as electric fencing are futile in some scenarios and high cost.
 
Cameras and computer vision are effective at identifying intruding animals. Some researchers~\cite{b3, b4, b5} use deep learning algorithms to recognize animals captured by the camera at regular intervals. However, fixed interval detection wastes resources and may miss some animals. Yadahalli~\emph{et.al.}~\cite{b6} instead send images to a TFT display and use a flash light for better night images, which are more expensive and consume more power. Compared to computer vision, it is also harder for humans to accurately identify animals in images where they make up a small percentage.

Cloud-based infrastructures~\cite{b10, b11, b12} are popular in smart agriculture for their powerful computing capabilities. In these systems, data from smart sensors is transmitted over the Internet to the cloud, where the data is stored and processed for decision making. However, these systems rely on Internet connectivity, which may be unavailable in rural areas, and can result in high latency due to data transmission to the cloud.

Many works~\cite{b14, b15, b16, b17, b18, b19, b20} that use infrared sensors lack specifics on sensor placement and algorithms. Some works either present their approach in a very generic way without details about the algorithm and adequate evaluation, or they fail to achieve better performance~\cite{b3}. Some works can not support large service coverage at a low cost~\cite{b2,b7,b8,b9}.
\vspace{-0.05in}
\section{Conclusion}
This paper outlines the design of a fog-based smart agriculture system that aims to transform traditional agriculture into a smart agriculture system. The system overcomes the challenges of high communication latency and Internet connectivity issues by incorporating fog computing and LoRa communication. It addresses the top concern of farmers, animal intrusion, by detecting and predicting the location of animals using low-cost PIR sensors and cameras. The paper also proposes three different sensor layouts and an algorithm. Finally, the three layouts are experimentally compared, and the effectiveness and accuracy of the algorithm are verified.

\end{document}